\documentclass[10pt]{article}
\usepackage{latexsym,graphicx}
\usepackage{latexsym,graphicx}
\newcommand{\be}{\begin{equation}}
\newcommand{\ee}{\end{equation}}
\def\n{\noindent}
\catcode `\@=11 \catcode `\@=12
\begin{document}
\begin{center}
\large{\bf {Thermodynamical Behaviour of Inhomogeneous Universe with Varying $\Lambda$ in Presence of
 Electromagnetic Field}} \\
\vspace{10mm}
\normalsize{Anil Kumar Yadav}\\
\vspace{4mm}
\textit{Department of Physics, Anand Engineering
College,Keetham, Agra-282 007, India} \\
\textit{E-mail: abanilyadav@yahoo.co.in, anilyadav.physics@gmail.com}\\
%\vspace{5mm}
%\normalsize{}
%\vspace{5mm}
%\normalsize{}
\end{center}
\vspace{10mm}
%\date{}
%\maketitle
\begin{abstract}
\n Thermodynamical Behaviour of Inhomogeneous Universe with Varying $ \Lambda $ in Presence of 
Electromagnetic Field is obtained. $ F_{12} $ is the non-vanishing component of 
electromagnetic field tensor. To get a deterministic solution, it is assumed that the 
free gravitational field is Petrov type-II non-degenerate.The value of cosmological constant is 
found to be small and pasitive supported by recent results from the supernovae observations recently obtained
by High-Z Supernovae Ia Team and Supernovae Cosmological Project. A relation between cosmological
constant and thermodynamical quantities is established. Some physical and geometric
properties of the model are also discussed.
\end{abstract}
\smallskip
\n Key words : Cosmology, Electromagnetic field, Inhomogeneous universe .\\
\n PACS: 98.80.Jk, 98.80.-k\\
%\newpage
%%%%%%%%%%%%%%%%%%%%%%%%%%%%%%%%%%%%%%%%%%%%%%%%%%%%%%%%%%%%%%%%%%%%%%%%%%%%%%%%%%%
%%%%%%%%%%%%%%%%%%%%%%%%%%%%%%%%% section 1 Introduction %%%%%%%%%%%%%%%%%%%%%%%%%%
\section{Introduction}
A lesson given by the history of cosmology is that the concept of 
the cosmological term revives in the days of crisis and we have more reasons
 than ever to belive that the cosmological term is the necessary ingredient
 of any cosmological model.The possibility of adding
a cosmological vacuum energy density to the Einstein field equations raises the question 
empirical justification of such a step. A positive cosmological constant helps
overcome the age problem, conected on the one side with the high estimates of the Hubble
parameters and with the age of the globular clusters on the other. Further, it seems that 
in order to retain the cold dark matter theory in the spacially flat universe most of the 
critical density should be provided by a passitive cosmological constant \cite{ref1,ref2}.
Observationnal data indicate that the cosmologial constant, if nonzero, is smaller than 
$ 10^{-55} ~  cm^-2 $. However, since everything that contributes to the vacuum energy acts 
as a cosmological constant it can not just be dropped without serious considerations.
Moreover particle physics expectations for $\Lambda$ exceeds its present value by the factor of 
order $ 10^{120} $ ie in a sharp contrast to observations.
To explain this apparent discrepancy the point of view has been adopted which allow the 
$\Lambda$ term to vary in time \cite{ref3}$-$\cite{ref11}. The idea of that during the evolution of 
universe the energy density of the vacuum decays into the particles thus leading to the decrease of 
the cosmological constant. As the result one has the creation of particles although the typical rate
of the creation is very small.

The creation of matter and entropy from vacuum has been studied via quantum field
theory in curved spacetime \cite{ref12,ref13}. Most cosmological models exhibit a 
singurarity which presents difficulties for interpreting quantum effects, because all 
macroscopic parameters of created particles are infinite there.This leads to the problem
of the initial vacuum.A regular vacuum for species of the created particles can be defined
in simple terms as a state where all mean values describing the particles, such as energy density,
number density, entropy etc are zero. But this simple condition is not achieved in many scenarios,
so that either one has to postulate an initial state beyond the singularity, or to assume that there 
was a non-zero number of particles at the initial vacuum. One attempt to overcome these problems is
via incorporating the effect of particle creation into Einstein's field equactions. In the present study 
I interpret the source of created particles as a decaying vacuum, described phenomenologically by a 
time-dependent cosmological constant $ \Lambda (t) $.        

There are significant observational evidence for the detection of Einstein’s
cosmological constant, $\Lambda$ or a component of material content of the universe that
varies slowly with time and space to act like $\Lambda $. Some of the recent discussions
on the cosmological constant “problem” and on cosmology with a time-varying
cosmological constant by Ratra and Peebles \cite{ref14}, and Sahni and Starobinsky
\cite{ref15}, point out that in the absence of any interaction with matter or radiation,
the cosmological constant remains a “constant”.However, in the presence of
interactions with matter or radiation, a solution of Einstein equations and the
assumed equation of covariant conservation of stress-energy with a time-varying
$\Lambda$ can be found. This entails that energy has to be conserved by a decrease in the
energy density of the vacuum component followed by a corresponding increase
in the energy density of matter or radiation (see also Carroll, Press and Turner
\cite{ref16}, Peebles \cite{ref17}, Padmanabhan \cite{ref18}).There is a plethora of astrophysical ev-
idence today, from supernovae measurements (Perlmutter et al. \cite{ref19}, Riess et
al. \cite{ref20}, Garnavich et al. \cite{ref21}, Schmidt et al. \cite{ref22}, Blakeslee et al. \cite{ref23}, Astier
et al. \cite{ref24}),the spectrum of fluctuations in the Cosmic Microwave Background
(CMB) \cite{ref25}, baryon oscillations \cite{ref26} and other astrophysical data, indicating
that the expansion of the universe is currently accelerating. The energy budget
of the universe seems to be dominated at the present epoch by a mysterious dark
energy component, but the precise nature of this energy is still unknown. Many
theoretical models provide possible explanations for the dark energy, ranging
from a cosmological term \cite{ref27} to super-horizon perturbations \cite{ref28} and time-
varying quintessence scenarios \cite{ref29}. These recent observations strongly favour
a significant and a positive value of $ \Lambda $ with magnitude
$\Lambda\left(\frac{G\hbar}{c^3}\right)\approx 10^{-123}$.

The standard Friedman-Robertson-Walker (FRW) cosmological model
prescribes a homogeneous and an isotropic distribution for its
matter in the description of the present state of the universe. At
the present state of evolution, the universe is spherically
symmetric and the matter distribution in the universe is on the
whole isotropic and homogeneous. But in early stages of evolution,
it could have not had such a smoothed picture. Close to the big bang
singularity, neither the assumption of spherical symmetry nor that
of isotropy can be strictly valid. So we consider plane-symmetric,
which is less restrictive than spherical symmetry and can provide an
avenue to study inhomogeneities. Inhomogeneous cosmological models
play an important role in understanding some essential features of
the universe such as the formation of galaxies during the early
stages of evolution and process of homogenization. The early
attempts at the construction of such models have done by Tolman
\cite{ref30} and Bondi \cite{ref31} who considered spherically
symmetric models. Inhomogeneous plane-symmetric models were
considered by Taub \cite{ref32,ref33} and later by Tomimura
\cite{ref34}, Szekeres \cite{ref35}. Recently,
Senovilla \cite{ref36} obtained a new class of exact solutions of
Einstein's equation without big bang singularity, representing a
cylindrically symmetric, inhomogeneous cosmological model filled
with perfect fluid which is smooth and regular everywhere satisfying
energy and causality conditions. Later, Ruis and Senovilla
\cite{ref37} have separated out a fairly large class of singularity
free models through a comprehensive study of general cylindrically
symmetric metric with separable function of $r$ and $t$ as metric
coefficients. Dadhich et al. \cite{ref38} have established a link
between the FRW model and the singularity free family by deducing
the latter through a natural and simple in-homogenization and
anisotropization of the former. Recently Bali and Tyagi\cite{ref39}, Pradhan et al\cite{ref40} 
obtained a plane-symmetric inhomogeneous cosmological
models of perfect fluid distribution with electro-magnetic field.

\par
The occurrence of magnetic fields on galactic scale is
well-established fact today, and their importance for a variety of
astrophysical phenomena is generally acknowledged as pointed out by
Zeldovich et al. \cite{ref41}. Also Harrison \cite{ref42} has
suggested that magnetic field could have a cosmological origin. As a
natural consequences, we should include magnetic fields in the
energy-momentum tensor of the early universe. The choice of
anisotropic cosmological models in Einstein system of field
equations leads to the cosmological models more general than
Robertson-Walker model \cite{ref43}. 
Strong magnetic fields can be created due to adiabatic compression
in clusters of galaxies. Primordial asymmetry of particle (say
electron) over antiparticle (say positron) have been well
established as C P (charged parity) violation. Asseo and Sol
\cite{ref44} speculated the large-scale inter galactic magnetic
field and is of primordial origin at present measure $10^{-8}$ G and
gives rise to a density of order $10^{-35} g cm^{-3}$. The present
day magnitude of magnetic energy is very small in comparison with
the estimated matter density, it might not have been negligible
during early stage of evolution of the universe. FRW models are
approximately valid as present day magnetic field is very small. The
existence of a primordial magnetic field is limited to Bianchi Types
I, II, III, $VI_{0}$ and $VII_{0}$ as shown by Hughston and Jacobs
\cite{ref45}. Large-scale magnetic fields give rise to anisotropies
in the universe. The anisotropic pressure created by the magnetic
fields dominates the evolution of the shear anisotropy and it decays
slower than if the pressure was isotropic \cite{ref46,ref47}. Such
fields can be generated at the end of an inflationary epoch
\cite{ref48}$-$\cite{ref50}. Anisotropic magnetic field models have
significant contribution in the evolution of galaxies and stellar
objects. 
%%%%%%%%%%%%%%%%%%%%%%%%%%%%%%%%%%%%%%%%%%%%%%%%%%%%%%%%%%%%%%%%%%%%%
%%%%%%%%%%%%%%%%%%%%%%%%%%%%%%%  SECTION 2  %%%%%%%%%%%%%%%%%%%%%%%%%%%%
\section{The metric and field  equations}
We consider the metric in the form of Marder \cite{ref51}
\begin{equation}
\label{eq1} ds^{2} = A^{2}(dx^{2} - dt^{2}) + B^{2} dy^{2} +
C^{2} dz^{2},
\end{equation}
where the metric potential $A$, $B$ and $C$ are functions of $x$ and $t$.
The energy momentum tensor is taken as
\begin{equation}
\label{eq2} T^{j}_{i} = (\rho + p)v_{i}v^{j} + p g^{j}_{i} + E^{j}_{i},
\end{equation}
where $E^{j}_{i}$ is the electro-magnetic field given by
Lichnerowicz \cite{ref52} as
\begin{equation}
\label{eq3} E^{j}_{i} = \bar{\mu}\left[h_{l}h^{l}(v_{i}v^{j} +
\frac{1}{2}g^{j}_{i}) - h_{i}h^{j}\right].
\end{equation}
Here $\rho$ and $p$ are the energy density and isotropic pressure respectively and
$v^{i}$ is the flow vector satisfying the relation
\begin{equation}
\label{eq4} g_{ij} v^{i}v^{j} = - 1.
\end{equation}
$\bar{\mu}$ is the magnetic permeability and $h_{i}$ the magnetic flux
vector defined by
\begin{equation}
\label{eq5} h_{i} = \frac{1}{\bar{\mu}}~~ ^*F_{ji}v^{j},
\end{equation}
where $^*F_{ij}$ is the dual electro-magnetic field tensor defined
by Synge \cite{ref53}
\begin{equation}
\label{eq6} ^*F_{ij} = \frac{\sqrt-g}{2}\epsilon_{ijkl} F^{kl}.
\end{equation}
$F_{ij}$ is the electro-magnetic field tensor and $\epsilon_{ijkl}$
is the Levi-Civita tensor density. The coordinates are considered to
be comoving so that $v^{1}$ = $0$ = $v^{2}$ = $v^{3}$ and $v^{4}$ =
$\frac{1}{A}$. We consider that the current is flowing along the
z-axis so that $h_{3} \ne 0$, $h_{1} = 0 = h_{2} = h_{4}$. The only
non-vanishing component of $F_{ij}$ is $F_{12}$. The Maxwell's
equations
\begin{equation}
\label{eq7} F_{ij;k} + F_{jk;i} + F_{ki;j} = 0,
\end{equation}
and
\begin{equation}
\label{eq8} \Biggl[\frac{1}{\bar{\mu}} F^{ij}\Biggr]_{;j} = 0,
\end{equation}
require that $F_{12}$ be function of $x$ alone. We assume that the magnetic
permeability as a function of $x$ and $t$ both. Here the semicolon
represents a covariant differentiation. \\

The Einstein's field equations ( in gravitational units c = 1, G
= 1 ) read as
\begin{equation}
\label{eq9} R^{j}_{i} - \frac{1}{2} R g^{j}_{i} + \Lambda
g^{j}_{i} = - 8\pi T^{j}_{i},
\end{equation}
for the line element (1) has been set up as
\[
8\pi A^{2}\left( p + \frac{F^{2}_{12}}{2\bar{\mu}A^{2}B^{2}}\right) = - \frac{B_{44}}
{B} - \frac{C_{44}}{C} + \frac{A_{4}}{A}\left(\frac{B_{4}}{B} + \frac{C_{4}}{C}\right)
\]
\begin{equation}
\label{eq10}
+ \frac{A_{1}}{A}\left(\frac{B_{1}}{B} + \frac{C_{1}}{C}\right) + \frac{B_{1}C_{1}}
{BC} - \frac{B_{4}C_{4}}{BC} - \Lambda A^{2},
\end{equation}
\begin{equation}
\label{eq11}
8\pi A^{2}\left( p + \frac{F^{2}_{12}}{2\bar{\mu}A^{2}B^{2}}\right) = -\left(\frac{A_{4}}
{A}\right)_{4} + \left(\frac{A_{1}}{A}\right)_{1} - \frac{C_{44}}{C} + \frac{C_{11}}{C}
- \Lambda A^{2},
\end{equation}
\begin{equation}
\label{eq12}
8\pi A^{2}\left(p - \frac{F^{2}_{12}}{2\bar{\mu}A^{2}B^{2}}\right) = -\left(\frac{A_{4}}
{A}\right)_{4} + \left(\frac{A_{1}}{A}\right)_{1} - \frac{B_{44}}{B} + \frac{B_{11}}{B}
- \Lambda A^{2},
\end{equation}
\[
8\pi A^{2}\left(\rho + \frac{F^{2}_{12}}{2\bar{\mu}A^{2}B^{2}}\right) = - \frac{B_{11}}
{B} - \frac{C_{11}}{C} + \frac{A_{1}}{A}\left(\frac{B_{1}}{B} + \frac{C_{1}}{C}\right)
\]
\begin{equation}
\label{eq13}
+ \frac{A_{4}}{A}\left(\frac{B_{4}}{B} + \frac{C_{4}}{C}\right) - \frac{B_{1}C_{1}}
{BC} + \frac{B_{4}C_{4}}{BC} + \Lambda A^{2},
\end{equation}
\begin{equation}
\label{eq14}
0 = \frac{B_{14}}{B} + \frac{C_{14}}{C} - \frac{A_{1}}{A}\left(\frac{B_{4}}{B} +
\frac{C_{4}}{C}\right) - \frac{A_{4}}{A}\left(\frac{B_{1}}{B} + \frac{C_{1}}{C}\right),
\end{equation}
where the sub indices $1$ and $4$ in A, B, C and elsewhere indicate ordinary
differentiation with respect to $x$ and $t$, respectively.
%%%%%%%%%%%%%%%%%%%%%%%%%%%%%%%%%%%%%%%%%%%%%%%%%%%%%%%%%%%%%%%%%%%%%
%%%%%%%%%%%%%%%%%%%%%%%%%%%%%%%  SECTION 3  %%%%%%%%%%%%%%%%%%%%%%%%%
\section{Solution of the field equations}
Equations (\ref{eq10}) - (\ref{eq12}) lead to
\[
\left(\frac{A_{4}}{A}\right)_{4} - \frac{B_{44}}{B} + \frac{A_{4}}{A}\left(\frac{B_{4}}{B}
+ \frac{C_{4}}{C}\right) - \frac{B_{4}C_{4}}{BC} =
\]
\begin{equation}
\label{eq15} \left(\frac{A_{1}}{A}\right)_{1} + \frac{C_{11}}{C} -
\frac{A_{1}}{A}\left(\frac{B_{1}}{B} + \frac{C_{1}}{C}\right) -
\frac{B_{1}C_{1}}{BC} = \mbox{a (constant)},
\end{equation}
and
\begin{equation}
\label{eq16}
 \frac{8\pi F^{2}_{12}}{\bar{\mu}B^{2}} = \frac{B_{44}}{B} - \frac{B_{11}}{B} +
\frac{C_{11}}{C} - \frac{C_{44}}{C}.
\end{equation}
Eqs. (\ref{eq10}) - (\ref{eq14}) represent a system of five
equations in seven unknowns $A$, $B$, $C$, $\rho$, $p$, $\Lambda$
and $\bar{\mu}$. For the complete determination of these unknowns
two more conditions are needed. As in the case of
general-relativistic cosmologies, the introduction of
inhomogeneities into the cosmological equations produces a
considerable increase in mathematical difficulty: non-linear partial
differential equations must now be solved. In practice, this means
that we must proceed either by means of approximations which render
the non-linearities tractable, or we must introduce particular
symmetries into the metric of the space-time in order to reduce the
number of degrees of freedom which the inhomogeneities can exploit.
In the present case, we assume that the metric is Petrov type-II
non-degenerate. This requires that
\[
\left(\frac{B_{11} + B_{44} + 2B_{14}}{B}\right) - \left(\frac{C_{11} + C_{4} + 2C_{14}}
{C}\right) =
\]
\begin{equation}
\label{eq17}
\frac{2(A_{1} + A_{4})(B_{1} + B_{4})}{AB} - \frac{2(A_{1} +
A_{4})(C_{1} + C_{4})}{AC}.
\end{equation}
Let us consider that
\[
A = f(x)\lambda(t),
\]
\[
B = g(x)\mu(t),
\]
\begin{equation}
\label{eq18}
C = g(x)\nu(t).
\end{equation}
Using (\ref{eq18}) in (\ref{eq14}) and (\ref{eq17}), we get
\begin{equation}
\label{eq19} \left[\frac{\frac{g_{4}}{g} -
\frac{f_{1}}{f}}{\frac{g_{1}}{g}}\right] =
\left[\frac{\frac{2\lambda_{4}}{\lambda}}{\frac{\mu_{4}}{\mu} +
\frac{\nu_{4}} {\nu}}\right] = \mbox{b (constant)},
\end{equation}
and
\begin{equation}
\label{eq20}
\frac{\frac{\mu_{44}}{\mu} - \frac{\nu_{44}}{\nu}}{\frac{\mu_{4}}{\mu} -
\frac{\nu_{4}}{\nu}} - \frac{2\lambda_{4}}{\lambda} = 2\left(\frac{f_{1}}{f} -
\frac{g_{1}}{g}\right) = \mbox{L (constant)}.
\end{equation}
Equation (\ref{eq19}) leads to
\begin{equation}
\label{eq21} f = ng^{(1 - b)},
\end{equation}
and
\begin{equation}
\label{eq22}
\lambda = m(\mu \nu)^{\frac{b}{2}},
\end{equation}
where m and n are constants of integration. Equations (\ref{eq15}), (\ref{eq18})
and (\ref{eq20}) lead to
\begin{equation}
\label{eq23} \left(\frac{b}{2} - 1\right)\frac{\mu_{44}}{\mu} + (b -
1)\frac{\mu_{4}\nu_{4}} {\mu \nu} = a,
\end{equation}
and
\begin{equation}
\label{eq24}
(2 - b)\frac{g_{11}}{g} + (3b - 4)\frac{g^{2}_{1}}{g^{2}} = a.
\end{equation}
Let us assume
\begin{equation}
\label{eq25} \mu = e^{U + W},
\end{equation}
and
\begin{equation}
\label{eq26}
\nu = e^{U - W}.
\end{equation}
Equations (\ref{eq20}), (\ref{eq25}) and (\ref{eq26}) lead to
\begin{equation}
\label{eq27}
W_{4} = Me^{Lt + 2(b - 1)U},
\end{equation}
where M is constant. From equations (\ref{eq23}), (\ref{eq25}), (\ref{eq26}) and
(\ref{eq27}), we have
\begin{equation}
\label{eq28}
(b - 1)U_{44} + 2(b - 1)U^{2}_{4} - 2bM e^{Lt + 2(b -1)U}U_{4} - ML e^{Lt + 2(b -1)U}
= a.
\end{equation}
If we put $e^{2U} = \xi$ in equation (\ref{eq28}), we obtain
\begin{equation}
\label{eq29} \frac{(b - 1)}{2}\frac{d^{2}\xi}{dt^{2}} - M
\frac{d}{dt}(e^{Lt}\xi^{b}) = a\xi.
\end{equation}
If we consider $\xi = e^{h t}$, then equation (\ref{eq29}) leads to
\begin{equation}
\label{eq30} \frac{(b - 1)}{2}g^{2}e^{h t} -
M\frac{d}{dt}(e^{Lt}e^{h b t}) = ae^{h t},
\end{equation}
which again reduces to
\begin{equation}
\label{eq31} h = \frac{L}{1 - b}
\end{equation}
and
\begin{equation}
\label{eq32}
a = \frac{L(L + 2M)}{2(b - 1)}.
\end{equation}
Thus
\begin{equation}
\label{eq33} U = \frac{Lt}{2(1 - b)}.
\end{equation}
Equations (\ref{eq27}) and (\ref{eq33}) reduce to
\begin{equation}
\label{eq34}
W = Mt + \log {N},
\end{equation}
where $N$ is an integrating constant. Eq. (\ref{eq24}) leads to
\begin{equation}
\label{eq35}
g = \beta \sinh^{\frac{2 -b}{2(b - 1)}}{(\alpha x + \delta)},
\end{equation}
where
$$
\alpha = \frac{\sqrt{2a(b-1)}}{(2 - b)}, ~~~~ \beta=\beta_{0}^{\frac{2 -b}{2(b - 1)}}
$$
and $\beta_{0}$, $\delta$ being constants of integration. Hence
\begin{equation}
\label{eq36}
f = n\beta \sinh^{\frac{b - 2}{2(b - 1)}}{(\alpha x + \delta)},
\end{equation}
\begin{equation}
\label{eq37}
\lambda = m e^{\frac{Lbt}{2(1 - b)}},
\end{equation}
\begin{equation}
\label{eq38}
\mu = e^{\frac{Lbt}{2(1 - b)} + Mt + \log{N}},
\end{equation}
\begin{equation}
\label{eq39}
\nu = e^{\frac{Lbt}{2(1 - b)} - Mt - \log{N}}.
\end{equation}
Therefore, we have
\begin{equation}
\label{eq40}
A = f \lambda =  m n \beta e^{\frac{Lbt}{2(1 - b)}}\sinh^{\frac{b -2}{2}}
{(\alpha x + \delta)},
\end{equation}
\begin{equation}
\label{eq41}
B = g \mu = N \beta e^{\left(\frac{L}{1 - b} + 2M \right)\frac{t}{2}}
\sinh^{\frac{2 - b}{2(b - 1)}}{(\alpha x + \delta)},
\end{equation}
\begin{equation}
\label{eq42}
C = g \nu = \frac{\beta}{N}e^{\left(\frac{L}{1 - b} - 2M \right)\frac{t}{2}}
\sinh^{\frac{2 - b}{2(b - 1)}}{(\alpha x + \delta)}.
\end{equation}
By using the transformation
\[
\alpha X = \alpha x + \delta,
\]
\[
Y = Gy,
\]
\[
Z = Hz,
\]
\begin{equation}
\label{eq43}
T = t,
\end{equation}
the metric (\ref{eq1}) reduces to the form
\[
ds^{2} = K^{2} \sinh^{b - 2}{(\alpha X)} e^{\frac{LTb}{1 - b}}(dX^{2} - dT^{2}) +
\]
\begin{equation}
\label{eq44}
\sinh^{\frac{2 - b}{b - 1}}{(\alpha X)} e^{\left(\frac{L}{1 - b}
+ 2M \right)T}dY^{2}   + \sinh^{\frac{2 - b}{b - 1}}{(\alpha X)}
e^{\left(\frac{L}{1 - b} - 2M\right)T}dZ^{2},
\end{equation}
where $K = mn\beta$, $G = N\beta$ and $H = \frac{\beta}{N}$.
%%%%%%%%%%%%%%%%%%%%%%%%%%%%%%%%%%%%%%%%%%%%%%%%%%%%%%%%%%%%%%%%%%%%%
%%%%%%%%%%%%%%%%%%%%%%%%%%%%%%%  SECTION 4  %%%%%%%%%%%%%%%%%%%%%%%%%
\section{Some Physical and Geometric Features}
The physical parameters, pressure $(p)$ and density $(\rho)$, for the model
(\ref{eq44}) are given by
$$
8\pi p = \frac{1}{K^{2}}e^{\frac{LbT}{b - 1}}\sinh^{2 - b}{(\alpha X)}\Biggl[
\frac{(2 - b)^{2} \alpha^{2}}{4(b - 1)}\left\{1 + \frac{2 - b}{b - 1}
\coth^{2}{(\alpha X)}\right\}
$$
\begin{equation}
\label{eq45}
 - \frac{L^{2}}{4(1 - b)^{2}} - M^{2}\Biggr] - \Lambda,
\end{equation}

$$
8\pi \rho = \frac{1}{K^{2}}e^{\frac{LbT}{b - 1}}\sinh^{2 - b}{(\alpha X)}\Biggl[
\frac{(2 - b) \alpha^{2}}{2(b - 1)}\left\{\frac{b}{b - 1}\coth^{2}{(\alpha X)}
- 1\right\}
$$
\begin{equation}
\label{eq46}
 + \frac{L^{2}(2b - 1)}{4(1 - b)^{2}} - M^{2} -\frac{ML}{(1 - b)}\Biggr] + \Lambda.
\end{equation}
In this case to find the explicit value of cosmological constant $\Lambda(t)$, one may assume that the fluid obey an 
equation of state of the form
\begin{equation}
\label{eq47}
 p=\gamma\rho
\end{equation}
where $\gamma(0\leq\gamma\leq1)$ is a constant.\\

Using equation (\ref{eq47}) in (\ref{eq45}) and then solving with(\ref{eq46}), we have
$$
8\pi (1+\gamma)\rho = \frac{1}{K^{2}}e^{\frac{LbT}{b - 1}}\sinh^{2 - b}{(\alpha X)}\Biggl[
\frac{(2 - b)(b^2 - 2b +4) \alpha^{2}}{4(b - 1)^{2}}\coth^{2}{(\alpha X)}
$$
\begin{equation}
\label{eq48}
+\frac{b(b-2)\alpha^2}{4(b-1)} - \frac{L^{2}}{2(1 - b)} - 2M^{2} -\frac{ML}{(1 - b)}\Biggr],
\end{equation}

\begin{figure}
\begin{center}
\includegraphics[width=4.0in]{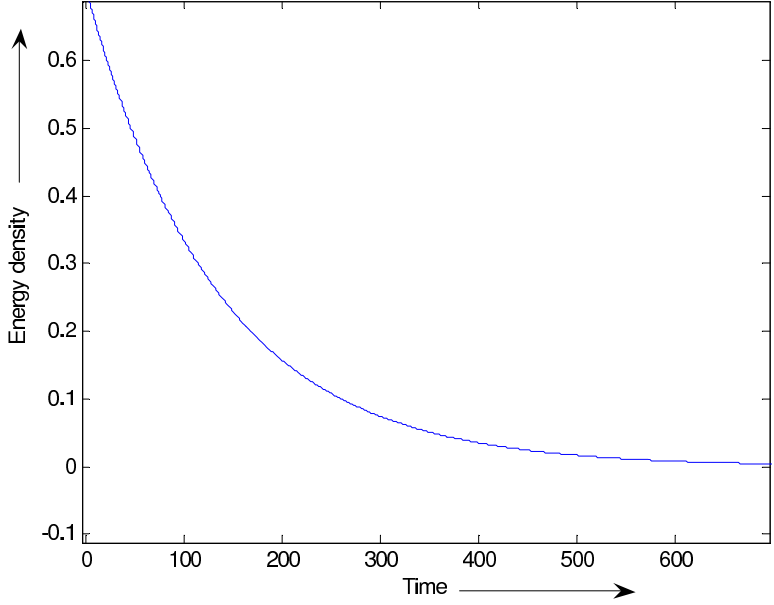} 
\caption{The plot of energy density $\rho$ vs. time T with parameters $ b=0.5, m = 0.03, 
 n = 0.25, K = 0.2 ~ and ~ \gamma = 0.4 $}
\label{fg:Fig1.eps}
\end{center}
\end{figure}

Eliminating $\gamma$ from (\ref{eq46}) and (\ref{eq48}), we obtain
$$
(1+\gamma)\Lambda = \frac{1}{K^{2}}e^{\frac{LbT}{b - 1}}\sinh^{2 - b}{(\alpha X)}\Biggl[
\frac{(2 - b)\left(b^2 - 4\gamma b +4(b+1)\right)\alpha^{2}}{4(b - 1)^{2}}\coth^{2}{(\alpha X)}
$$
\begin{equation}
\label{eq49}
-\frac{(b-2)(2+2\gamma-b)\alpha^2}{4(b-1)} - \frac{L^{2}\left(\gamma(2b-1)+1\right)}{4(1 - b)^2} - (1-\gamma)M^{2} 
+\frac{\gamma ML}{(1 - b)}\Biggr],
\end{equation}

\begin{figure}
\begin{center}
\includegraphics[width=4.0in]{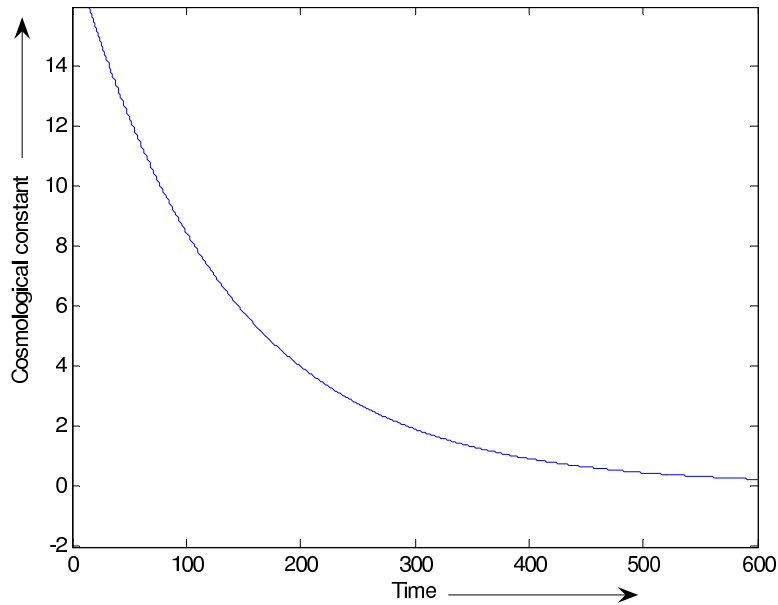} 
\caption{The plot of cosmological constant $\Lambda$ vs. time T with parameters $ b=0.5, m = 0.03, 
 n = 0.25, K = 0.2 ~ and ~ \gamma = 0.4 $}
\label{fg:Fig2.eps}
\end{center}
\end{figure}

From (\ref{eq46}) and (\ref{eq48}), we note that $\rho(t)$ is the decreasing function of time and $\rho > 0$ for 
all time. This behaviour is clearly shown in figure 1, as a representative case with appropriate choice of constants
of integration and other physical parameters using reasonably well known situations.

In spite of homogeniety at large scale our universe is inhomogenrous at small scale, so physical quantities
being position dependent are more natural in our observable universe if we do not go to super high scale.
This result show this kind of physical importance. In recent time theoreticians and observers draw high attention 
on $\Lambda$ - terms due to various reasons.The non-trivial role of the vacuum in early universe generates a 
$\Lambda$ that leads to inflationary phase. Obsevationally this term provides an additional parameter to accommodate
conflicting data on the value of Hubble's constant, deceleration parameter, the density parameters and the age of 
universe \cite{ref54} and \cite{ref55}. Assuming that $\Lambda$ owes it origin to vacuum interactions, as suggested
in particular by Sakharav \cite{ref56}, it follows that it would in general be a function of space and 
time co-ordinates, rather than a strict constant. In a homogeneous universe $\Lambda$ will be most time dependent
 \cite{ref57}. In our case this approach can generate $\Lambda$ that varies both with space and time. In considering 
the nature of local massive objects, however the space dependence of $\Lambda$ can not ignored. For detail discussion,
the readers are advised to see the reference (Narlikar, Pecker and Vigier \cite{ref58}, Ray and Ray \cite{ref59},
Tiwari, Ray and Bhadra \cite{ref60}). 

From (\ref{eq49}), we see that cosmological constant $\Lambda$ is a decreasing function of time and it approaches
a small positive value at late time. This behaviour is clearly shown in figure 2. Recent cosmological obsevations
suggest the existence of positive cosmological constant $\Lambda$ with the magnitude 
$\Lambda\left(\frac{G\hbar}{c^3}\right)\approx 10^{-123}$. These observations on magnitude and red-shift of 
type Ia supernova suggest that our universe may be an accelerating one with induced 
cosmological density through the cosmological
$\Lambda$-term. Thus the model presented in this paper is consistent with the results of recent observations.\\ 
The non-vanishing component $F_{12}$ of the electromagnetic field tensor is
given by
\begin{equation}
\label{eq50}
F_{12} = \sqrt{\frac{\bar{\mu}}{8\pi}\frac{2ML}{(1 - b)}} G e^{\left(\frac{L}{1 - b}
+ 2M \right)\frac{T}{2}}\sinh^{\frac{2 - b}{2(b - 1)}}{(\alpha X)},
\end{equation}
where $\bar{\mu}$ remains undetermined as function of $x$ and $t$ both. \\
The scalar of expansion $(\theta)$ calculated for the flow vector $(v^{i})$ is given by
\begin{equation}
\label{eq51} \theta = \frac{L(b + 2)}{2K(1 - b)} e^{\frac{LbT}{2(b -
1)}}\sinh^{\frac{(2 - b)}{2}} {(\alpha X)}.
\end{equation}
The shear scalar $(\sigma^{2})$, acceleration vector
$(\dot{v}_{i})$, deceleration parameter $q$, proper volume $V$
and Hubble parameter $H$ for the model (\ref{eq45}) are given by
\begin{equation}
\label{eq52}
\sigma^{2} = \frac{(L^{2} + 12 M^{2})}{12 K^{2}}e^{\frac{LbT}{(b - 1)}}
\sinh^{(2 - b)}{(\alpha X)},
\end{equation}
\begin{equation}
\label{eq53}
\dot{v}_{i} = \left(\frac{1}{2}(b - 2)\alpha \coth(\alpha X), 0, 0, 0\right),
\end{equation}
\begin{equation}
\label{eq54} q = - \frac{4K^{2}(b + 1)^{2}}{81(b +
2)^{2}}\exp{\left(\frac{L b T}{1 - b}\right)}\sinh^{(b - 2)}{(\alpha
X)},
\end{equation}
\begin{equation}
\label{eq55} V = \sqrt{-g} = K^{2} G H e^{\frac{L(b + 1)T}{(1 -
b)}}\sinh^{\frac{(b - 2)(b - 3)}{(b - 1)}} {(\alpha X)},
\end{equation}
\begin{equation}
\label{eq56} H = \frac{3L(b + 2)}{2K(1 - b)} e^{\frac{LbT}{2(b -
1)}}\sinh^{\frac{(2 - b)}{2}} {(\alpha X)}.
\end{equation}
From equations(\ref{eq51}) and (\ref{eq52}), we have
\begin{equation}
\label{eq57} \frac{\sigma^{2}}{\theta^{2}} = \frac{(L^{2} + 12
M^{2})(1 - b^{2})}{3L^{2}(b + 2)^{2}} = \mbox{constant}.
\end{equation}
The rotation $\omega$ is identically zero and the non-vanishing component of
conformal curvature tensor are given by
\begin{equation}
\label{eq58} C_{(1212)} = \frac{1}{6K^{2}}e^{\frac{LbT}{(b -
1)}}\sinh^{(2 - b)}{(\alpha X)} \left[b \alpha - \frac{L^{2}}{4b} +
3ML - 2M^{2}\right],
\end{equation}
\begin{equation}
\label{eq59} C_{(1313)} = \frac{1}{6K^{2}}e^{\frac{LbT}{(b -
1)}}\sinh^{(2 - b)}{(\alpha X)} \left[b \alpha - \frac{L^{2}}{4b} -
3ML - 2M^{2}\right],
\end{equation}
\begin{equation}
\label{eq60} C_{(2323)} = \frac{1}{3K^{2}}e^{\frac{LbT}{(b -
1)}}\sinh^{(2 - b)}{(\alpha X)} \left[- b\alpha + \frac{L^{2}}{4b} +
\frac{ML}{(1 - b)} + 2M^{2}\right],
\end{equation}
\begin{equation}
\label{eq61} C_{(1224)} =\frac{ML}{2K^{2}}e^{\frac{LbT}{(b -
1)}}\sinh^{(2 - b)} {(\alpha X)}.
\end{equation}
The reality conditions (Ellis \cite{ref61})
$$
(i) \rho + p > 0, ~ ~ (ii) \rho + 3p > 0,
$$
lead to
\[
\frac{(2 - b)\alpha^{2}}{4(b - 1)^{2}}\left[(b^{2} - 2b + 4)\coth^{2}{(\alpha X)}
+ b(1 - b)\right] >
\]
\begin{equation}
\label{eq60} \frac{L^{2}}{2(1 - b)} + 2 M^{2} + \frac{ML}{(1 - b)},
\end{equation}
and
\[
e^{\frac{LbT}{(b - 1)}}\Biggl[\frac{(2 - b)(3b^2 -10b +12)\alpha^{2}}{4(b - 1)^2}\coth^{2}{(\alpha X)} + 
\frac{(2-b)(3b-4)\alpha^2}{4(1-b)}
\]
\begin{equation}
\label{eq61} + \frac{L^{2}(2b-4)}{4(1 - b)^{2}} - \frac{ML}{(1 - b)} -4M^2
\Biggr] + 2K^{2}\Lambda \sinh^{(b - 2)}{(\alpha X)} > 0
\end{equation}
respectively.\\
The dominant energy condition is given by Hawking and Ellis
\cite{ref62}
$$
(i)  \rho - p \geq 0, ~ ~ (ii) ~ ~ ~ ~ \rho + p \geq 0
$$
lead to
\[
e^{\frac{LbT}{(b - 1)}}\Biggl[\frac{(2 - b)\alpha^{2}}{2(b - 1)}\left\{\frac{(b^{2}
- 6b +4)}{2(1 - b)}\coth^{2}{(\alpha X)} + \frac{b - 4}{2}\right\}
\]
\begin{equation}
\label{eq61} + \frac{L^{2}b}{2(1 - b)^{2}} - \frac{ML}{(1 - b)}
\Biggr] + 2K^{2}\Lambda \sinh^{(b - 2)}{(\alpha X)} \geq 0
\end{equation}
and
\[
\frac{(2 - b)\alpha^{2}}{4(b - 1)^{2}}\left[(b^{2} - 2b + 4)\coth^{2}{(\alpha X)}
+ b(1 - b)\right] \geq
\]
\begin{equation}
\label{eq62} \frac{L^{2}}{2(1 - b)} + 2 M^{2} + \frac{ML}{(1 - b)},
\end{equation}
respectively.
%%%%%%%%%%%%%%%%%%%%%%%%%%%%%%%%%%%%%%%%%%%%%%%%%%%%%%%%%%%%%%%%%%%%%%%%%%%%%%%
%%%%%%%%%%%%%%%%%%%%%%%%%%% SECTION 5 %%%%%%%%%%%%%%%%%%%%%%%%%%%%%%%%%%%%%%%%
\section{Thermodynamical behaviour and entropy of universe}
From the thermodynamics \cite{ref63,ref64}, we apply the combination of first and second law of thermodynamics
to the system with volume V. As we know that
\begin{equation}
 \label{eq63}
 \c{T} dS=d(\rho V)+pdV
\end{equation}
where $ \c{T} $, S represents the tempreture and entropy respectively. \\
Eq. (\ref{eq63}) may be written as
\begin{equation}
 \label{eq64}
 \c{T} dS=d{\left[(\rho + p)V\right]} - Vdp
\end{equation}
The integrability condition is necessary to define a perefect fluid as a thermodynamical 
syetem \cite{ref65,ref66}. It is given by
\begin{equation}
 \label{eq65}
 dp=\frac{\rho + P}{\c{T}}d\c{T}
\end{equation}
Plugging eq. (\ref{eq65}) in eq. (\ref{eq64}), we have the differential equation
\begin{equation}
 \label{eq66}
 dS=\frac{1}{\c{T}}d{\left[(\rho + p)V\right]} - (\rho + p)V\frac{d\c{T}}{\c{T}^2}
\end{equation}
we rewrite eq (\ref{eq66}) as
\begin{equation}
 \label{eq67}
 dS=d{\left[\frac{(\rho + p)V}{\c{T}} + c\right]}
\end{equation}
where c is constant.\\
Hence the entropy is defined as
\begin{equation}
 \label{68}
 S=\frac{\rho + p}{\c{T}} V
\end{equation}
Let the entropy density be s, so that
\begin{equation}
 \label{eq69}
 s=\frac{S}{V}=\frac{\rho + p}{\c{T}}=\frac{(1+\gamma)\rho}{\c{T}}
\end{equation}
where $ p=\gamma\rho $ and $ 0<\gamma\leq1 $.\\
If we define the entropy density in terms of temprature then the first law of thermodynamics may be written as
\begin{equation}
 \label{eq70}
 d(\rho V)+\gamma\rho dV=(1+\rho)\c{T}d\left(\frac{\rho V}{\c{T}}\right)
\end{equation}
which on integration yields
\begin{equation}
 \label{71}
 \c{T}= c_{0}\rho^{\frac{\gamma}{(1+\gamma)}}
\end{equation}
where $ c_{0} $ is constant of integration.\\

From eqs. (\ref{eq69}) and (71), we obtain
\begin{equation}
 \label{eq72}
 s=\left(\frac{1+\gamma}{c_{0}}\right)\rho^{\frac{1}{1+\rho}}
\end{equation}
These equation are not valid for $ \gamma=-1 $. For the Zel'dovich fluid $ (\gamma=1) $, we get
\begin{equation}
 \label{eq73}
 \c{T}=c_{0}\rho^\frac{1}{2}
\end{equation}
\begin{equation}
 \label{eq74}
 s=\frac{2}{c_{0}}\rho^\frac{1}{2}
\end{equation}
 $$ \Rightarrow s\sim \rho^\frac{1}{2}\sim \c{T} $$ 
Thus the entropy density is proportional to the tempreture.\\ 
Now the tempreture, entropy density and entropy of Zel'dovich universe is given by
$$
\c{T} = T_{0}\Biggl[\frac{1}{K^{2}}e^{\frac{LbT}{b - 1}}\sinh^{2 - b}{(\alpha X)}\Biggl[
\frac{(2 - b) \alpha^{2}}{2(b - 1)}\left\{\frac{b}{b - 1}\coth^{2}{(\alpha X)}
- 1\right\}
$$
\begin{equation}
\label{eq76}
 + \frac{L^{2}(2b - 1)}{4(1 - b)^{2}} - M^{2} -\frac{ML}{(1 - b)}\Biggr] + \Lambda\Biggl]^\frac{1}{2}.
\end{equation}
$$
s = s_{0}\Biggl[\frac{1}{K^{2}}e^{\frac{LbT}{b - 1}}\sinh^{2 - b}{(\alpha X)}\Biggl[
\frac{(2 - b) \alpha^{2}}{2(b - 1)}\left\{\frac{b}{b - 1}\coth^{2}{(\alpha X)}
- 1\right\}
$$
\begin{equation}
\label{eq77}
 + \frac{L^{2}(2b - 1)}{4(1 - b)^{2}} - M^{2} -\frac{ML}{(1 - b)}\Biggr] + \Lambda\Biggl]^\frac{1}{2}.
\end{equation}
$$
S = S_{0}\Biggl[e^{\frac{(b-2)LT}{b - 1}}\sinh^{\frac{(b-2)(2b-7)}{(b-1)}}{(\alpha X)}\Biggl[
\frac{(2 - b) \alpha^{2}}{2(b - 1)}\left\{\frac{b}{b - 1}\coth^{2}{(\alpha X)}
- 1\right\}
$$
\begin{equation}
\label{eq78}
 + \frac{L^{2}(2b - 1)}{4(1 - b)^{2}} - M^{2} -\frac{ML}{(1 - b)}\Biggr] + \Lambda\Biggl]^\frac{1}{2}.
\end{equation}
where $ T_{0}=\frac{c_{0}}{\sqrt{8\pi}} $, $ s_{0}=\frac{2}{c_{0}\sqrt{8\pi}} $ and $ S_{0}=s_{0}K $ are constant.\\
For radiating fluid $ (\gamma=\frac{1}{3}) $, we get
\begin{equation}
 \label{eq79}
 \c{T}\sim\rho^\frac{1}{4}
\end{equation}
\begin{equation}
 \label{eq80}
 s\sim\rho^\frac{3}{4}\sim\c{T}^3
\end{equation} 
Thus the entropy density is proportional to cube of tempreture.\\
Now the tempreture, entropy density and entropy of radiating universe is given by
$$
\c{T} = T_{00}\Biggl[\frac{1}{K^{2}}e^{\frac{LbT}{b - 1}}\sinh^{2 - b}{(\alpha X)}\Biggl[
\frac{(2 - b) \alpha^{2}}{2(b - 1)}\left\{\frac{b}{b - 1}\coth^{2}{(\alpha X)}
- 1\right\}
$$
\begin{equation}
\label{eq81}
 + \frac{L^{2}(2b - 1)}{4(1 - b)^{2}} - M^{2} -\frac{ML}{(1 - b)}\Biggr] + \Lambda\Biggl]^\frac{1}{4}.
\end{equation}
$$
s = s_{00}\Biggl[\frac{1}{K^{2}}e^{\frac{LbT}{b - 1}}\sinh^{2 - b}{(\alpha X)}\Biggl[
\frac{(2 - b) \alpha^{2}}{2(b - 1)}\left\{\frac{b}{b - 1}\coth^{2}{(\alpha X)}
- 1\right\}
$$
\begin{equation}
\label{eq82}
 + \frac{L^{2}(2b - 1)}{4(1 - b)^{2}} - M^{2} -\frac{ML}{(1 - b)}\Biggr] + \Lambda\Biggl]^\frac{3}{4}.
\end{equation}
$$
S = S_{00} e^{\frac{L(b + 1)T}{(1 -b)}}\sinh^{\frac{(b - 2)(b - 3)}{(b - 1)}} {(\alpha X)}
\Biggl[\frac{1}{K^{2}}e^{\frac{LbT}{b - 1}}\sinh^{2 - b}{(\alpha X)}
$$
\begin{equation}
\label{eq82}\Biggl[
\frac{(2 - b) \alpha^{2}}{2(b - 1)}\left\{\frac{b}{b - 1}\coth^{2}{(\alpha X)}
- 1\right\}
 + \frac{L^{2}(2b - 1)}{4(1 - b)^{2}} - M^{2} -\frac{ML}{(1 - b)}\Biggr] + \Lambda\Biggl]^\frac{3}{4}.
\end{equation}
where $ T_{00}=\frac{c_{0}}{(8\pi)^{1/2}} $, $ s_{00}=\frac{2}{c_{0}(8\pi)^{1/2}} $ and 
$ S_{00}=s_{0}K^2 $ are constant.\\
%%%%%%%%%%%%%%%%%%%%%%%%%%%%%%%%%%%%%%%%%%%%%%%%%%%%%%%%%%%%%%%%%%%%%%%%%%%%%%%%
%%%%%%%%%%%%%%%%%%%%%%%%% SECTION 6 %%%%%%%%%%%%%%%%%%%%%%%%%%%%%%%%%%%%%%%%%%%%
\section{Discussion and Concluding Remarks }
The present study deals the plane-symmetric inhomogeneous cosmological
model of electro-magnetic perfect fluid as the source of matter. FRW
models are appoximatly valid, as present day magnetic field is very
small. Maarteens \cite{ref67} in his study explained that magnetic
fields are observed not only in stars but also in galaxies. In
princple, these fields could play a significant role in structure
formation but also affect the anisotropies in cosmic microwave
background radiation [CMB]. Since the electric and magnetic fields
are interrelated, their independent nature disappears when we
consider them as time dependance. Hence, it would be proper to look
upon these fields as a single field - electromagnetic field.
Generally the model represents expanding, shearing, non-rotating and
Petrov type-II non-degenerate universe in which the flow vector is
geodesic. We find that the model starts expanding at $T = 0$ and
goes on expanding indefinitely. For
large values of $T$, the model is conformally flat and Petrov
type-II non-degenerate. Since $\frac{\sigma}{\theta} = $
constant, hence the model does not approach isotropy.
The value of cosmological constant is found to be small and pasitive
supported by recent results from the supernovae observations recently obtained
by High-Z Supernovae Ia Team and Supernovae Cosmological Project.The relation 
between cosmological constant and thermodynamical quantities is 
established and from eqs. (\ref{eq80}) and $(85)$,
it is clear that cosmological constant affect entropy.  
The electromagnetic field tensor does not vanish when $L \ne 0$, $M \ne
0$, and $b \ne 1$. For large values of $T$ and $L + 2M(1 - b) < 0$,
$F_{12}$ tends to zero.
For $ b = 1$, we obtain singularity and model approach isotropy. $b
< 1$ and $b > 1$ imposed the restriction on the value of $M$ and $L$
which affect all the physical and
kinematical parameters of the model. \\

In spite of homogeneity at large scale our universe is
inhomogeneous at small scale, so physical quantities being
position-dependent are more natural in our observable universe if we
do not go to super high scale. Our derived model shows this kind of
physical importance. The expressions for deceleration parameter
($q$) and Hubble parameter ($H$) given by Eqs. (\ref{eq54}) and
(\ref{eq56}) respectively are functions of x and t as in the case of
inhomogeneous cosmological models. Obviously these two expressions
are different from that of FRW cosmological model.
The present study also extend the work of Yadav and Bagora \cite{ref68}  
 for inhomogeneous universe with varying $ \Lambda $ and clarify thermodynamics 
of inhomogeneous universe by introducing the integrability condition and 
tempreture. It is found that $s\sim \rho^\frac{1}{2}\sim \c{T}$ and 
$ s\sim\rho^\frac{3}{4}\sim\c{T}^3 $, for Zeldo'vich and radiating fluid model respectively.
From these expressions it is clear that $ \rho $ is the function of tempreture and volume, thus 
a new general equation of state describing the Zel'dovich fluid and 
radiating fluid models as a function of tempreture and volume is found. 
The basic equations of thermodynamics for inhomogeneous universe has been deduced
which may be useful for better understanding of evolution of universe.   

%\newline
%\nonumsection{References}

\end{document}